\documentclass[aps,prl,superscriptaddress,twocolumn,showpacs]{revtex4}

\usepackage{epsfig}

\newcommand{\tfrac}[2]{{\textstyle{\frac{#1}{#2}}}}
\def\msb{$\overline{\rm MS} $ }
\def\be{\begin{equation}}
\def\ee{\end{equation}}
\def\ba{\begin{eqnarray}}
\def\ea{\end{eqnarray}}

\begin{document}

\title{Charm and Bottom Quark Masses from Perturbative QCD}

\author{R. Boughezal}

\affiliation{Institut f\"ur Theoretische Physik und Astrophysik,
    Universit\"at W\"urzburg, \\ Am Hubland, D-97074 W\"urzburg,
    Germany}

\author{M. Czakon}

\affiliation{Institut f\"ur Theoretische Physik und Astrophysik,
  Universit\"at W\"urzburg, \\ Am Hubland, D-97074 W\"urzburg,
  Germany}

\affiliation{Department of Field Theory and Particle Physics,
  Institute of Physics, \\ University of Silesia, Uniwersytecka 4,
  PL-40007 Katowice, Poland}

\author{T. Schutzmeier}

\affiliation{Institut f\"ur Theoretische Physik und Astrophysik,
  Universit\"at W\"urzburg, \\ Am Hubland, D-97074 W\"urzburg,
  Germany}

\begin{abstract}

Using a new result for the first moment of the hadronic production cross
section at order ${\cal O}(\alpha_s^3)$, and new data on the $J/\psi$
and $\psi'$ resonances for the charm quark, we determine the \msb masses of the charm and
bottom quarks to be $\overline{m}_c(\overline{m}_c) = 1.295 \pm 0.015$
GeV and $\overline{m}_b(\overline{m}_b) = 4.205 \pm 0.058$ GeV. 
We assume that the continuum contribution to the sum rules is
adequately described by pQCD. While
we observe a large reduction of the perturbative error, the shifts
induced by the theoretical input are very small. The main change in
the central value of $m_c$ is related to the experimental data. On the
other hand, the value of $m_b$ is not changed by our calculation to
the assumed precision.
\end{abstract}

\pacs{12.38.-t, 14.65.Dw, 14.65.Fy}

\maketitle

The strong coupling constant, $\alpha_s$, and quark masses are the
only input parameters of the QCD lagrangian. Since at high energy
masses become essentially negligible, $\alpha_s$ can be determined
separately to a good precision. On the other hand, the determination
of the masses is a much more cumbersome problem, where the details of
the interaction can hardly be overcome. It is, therefore, not
surprising that it has attracted a lot of attention. Particularly
interesting are the charm and bottom quarks, where results can be
derived from lattice, see {\it e.g.} \cite{Dougall:2005ev}, or from
different types of sum rules, see {\it e.g.}
\cite{Penarrocha:2001ig,Eidemuller:2000rc,Hoang:2004wh}. In this
Letter, we shall follow the approach of \cite{Kuhn:2001dm}, which
requires, as input on the theoretical side, moments of the hadronic
cross section. The first moment at ${\cal O}(\alpha_s^3)$ constitutes
our main new result \footnote{During the preparation of the present
publication, Ref.~\cite{Chetyrkin:2006xg} appeared, which also
contains the first moment of the hadronic cross section. We find
perfect agreement between the two results.}, which we use, together
with new resonance data for $J/\psi$ and $\psi'$
\cite{Eidelman:2004wy}, to derive the most up to date values of
$\overline{m}_c(\overline{m}_c)$ and $\overline{m}_b(\overline{m}_b)$
in the \msb scheme.

The use of QCD sum rules for the determination of the charm quark mass
has originally been proposed in a series of papers gathered in
\cite{Novikov:1977dq}. The same approach has subsequently been applied
to the bottom quark \cite{Reinders:1984sr}. A specific feature of
these early studies has been the use of higher moments of the hadronic
cross section. The advantages of the lower moments have been studied
in \cite{Shifman:1978bx}, and used at order ${\cal O}(\alpha_s^2)$ in
\cite{Kuhn:2001dm} and subsequently in
\cite{Corcella:2002uu,Hoang:2004xm}. The latter analysis was based on
the calculation of three-loop moments from \cite{Chetyrkin:1995ii} and
the data of the cross section scan around the charm threshold given in
\cite{Bai:2001ct}. An important conclusion of \cite{Kuhn:2001dm} is
that the first moment is best suited for the analysis, since it has
the weakest dependence on the nonperturbative effects and details of
the threshold region, leading naturally to the smallest error.

\begin{figure}
\epsfig{file=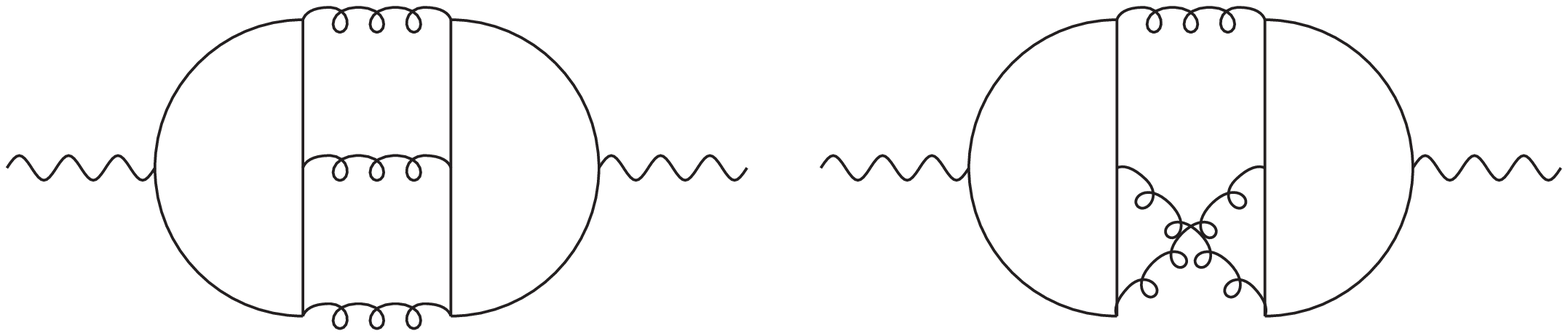,width=8cm}
\caption{\label{singlet} Singlet diagrams with a three-gluon cut.}
\end{figure}

Starting from ${\cal O}(\alpha_s^3)$, the perturbative cross section
receives contributions from diagrams with a massless threshold,
Fig.~\ref{singlet}. Fortunately, gauge invariance restrictions on the
photon-three-gluon vertex allow such diagrams only at ${\cal
O}((q^2/m^2)^4 \log(q^2/m^2))$
\cite{Groote:2001py,Portoles:2002rt}. Since we will use the first
moment exclusively, we can safely neglect them in our analysis and
restrict the calculation to diagrams where a single heavy quark line
connects both electromagnetic current vertices.

Let us now remind a few definitions. The hadronic ratio is given by
\be R(s) = \frac{\sigma(e^+ e^- \rightarrow {\rm had.})}{\sigma(e^+
e^- \rightarrow \mu^+ \mu^-)}, \ee where $\sigma(e^+ e^- \rightarrow
\mu^+ \mu^-) = 4\pi\alpha^2/3s$. It can be related to the
current-current correlator through \be R(s) = 12 \pi {\rm Im} \Pi(q^2
= s+i\epsilon), \ee where \be (-q^2g_{\mu\nu}+q_\mu q_\nu) \Pi(q^2) =
i \int dx e^{iqx} \langle 0 | T j_\mu(x) j_\nu(0)| 0 \rangle.  \ee For
a given heavy quark of mass $m$, the contribution to the correlator
can be expanded as \be
\label{moments}
\Pi_h(q^2) = Q_h^2 \frac{3}{16\pi^2} \sum_{n>0} \overline{C}_n z^n,
\;\;\;\;
z=\frac{q^2}{4m^2(\mu)}.
\ee
We perform all our calculations in the \msb scheme, and expand the
$\overline{C}_n$ coefficients in the strong coupling as
\be
\overline{C}_n = \sum_{k \ge 0}
\left(\frac{\alpha_s(\mu)}{\pi}\right)^k
\overline{C}_n^{(k)}\left(\log\left( \frac{m^2(\mu)}{\mu^2}\right)\right) .
\ee

The coefficients of the expansion in the strong coupling constant are
known exactly up to two-loops 
and all moments can be derived from a generating formula
\cite{Broadhurst:1993mw,barbieri}. The first eight moments of the
three-loop result have been given in \cite{Chetyrkin:1995ii}. Our new
result is the four-loop correction to the first moment, {\it i.e.} the
coefficient $\overline{C}_1^{(3)}$. We have obtained it by a direct
Taylor expansion to second order in $q^2$ of the photon self-energy,
followed by a reduction of the resulting four-loop single scale
tadpoles using an approach inspired by the Laporta algorithm
\cite{Laporta:2001dd}, see also \cite{Czakon:2004bu}. We have kept a single power of the linear gauge 
parameter in the 0th moment (value of $\Pi(0)$) to check for gauge
invariance, but neglected it in the 1st moment to spare unnecessary
computational complexity.

\begin{figure}
\epsfig{file=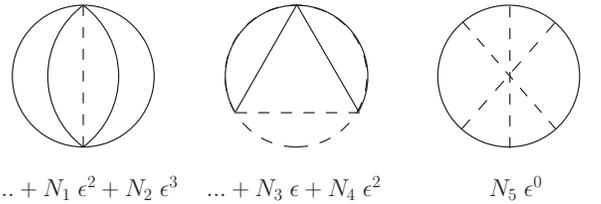,width=8cm}
\caption{\label{masters} Definition of the numerical constants $N_i$
  from the $\epsilon$ expansion of vacuum integrals.}
\end{figure}

The purely bosonic contribution to $\overline{\rm C}_1^{(3)}$ is given by
\begin{widetext}
\ba
&+& C_F^3  \,  \biggl( \tfrac{45577758023909}{56435097600} - \tfrac{454}{2835} \, N_5 + \tfrac{173}{1080} \, N_4 + 
\tfrac{1078129}{3628800} \, N_3 + \tfrac{79}{1920} \, N_2 - \tfrac{8881}{604800} \, N_1 - \tfrac{3546972523}{4702924800} \, \pi^2 
+ \tfrac{1533371173}{653184000} \, \pi^4 \nonumber \\ &&+ \tfrac{24802703}{342921600} \, \pi^6
+ \tfrac{8192}{45} \, a_5 - \tfrac{252157}{4050} \, a_4 + \tfrac{79}{3360} \, \zeta_7 - \tfrac{5193373091}{27216000} \, \zeta_5
+ \tfrac{79}{7200} \, \zeta_5 \, \pi^2 - \tfrac{75721242853}{195955200} \, \zeta_3 - \tfrac{95128427}{48988800} \, \zeta_3 \, 
\pi^2  \nonumber \\ && + \tfrac{79}{21600} \, \zeta_3 \, \pi^4 - \tfrac{10306609}{4082400} \, \zeta_3^2 + \tfrac{4352}{2025} \, l_2 \, \pi^4
+ \tfrac{252157}{97200} \, l_2^2 \, \pi^2 + \tfrac{1024}{405} \, l_2^3 \, \pi^2 - \tfrac{252157}{97200} \, l_2^4 - 
\tfrac{1024}{675} \, l_2^5 - \tfrac{6142460327}{22394880} \, l_m - \tfrac{173}{270} \, l_m \, N_3 \nonumber \\ && - \tfrac{79}{480} \, l_m \, N_1
+ \tfrac{12907273}{1866240} \, l_m \, \pi^2 + \tfrac{215987}{777600} \, l_m \, \pi^4 + \tfrac{79}{45360} \, l_m \, 
\pi^6 - \tfrac{16384}{135} \, l_m \, a_4 + \tfrac{1962691}{32400} \, l_m \, \zeta_5 + \tfrac{6307081}{155520} \, l_m \, \zeta_3 \nonumber \\
&& + \tfrac{36163}{58320} \, l_m \, \zeta_3 \, \pi^2 + \tfrac{79}{540} \, l_m \, \zeta_3^2 + \tfrac{2048}{405} \, l_m \, l_2^2 \, \pi^2 - 
\tfrac{2048}{405} \, l_m \, l_2^4 - \tfrac{103}{90} \, l_m^2 + \tfrac{3}{5} \, l_m^3 \biggr)
\nonumber \\
&+& C_F^2 \, C_A  \,  \biggl(  - \tfrac{4622898495103}{5374771200} + \tfrac{227}{945} \, N_5 - \tfrac{103}{288} \, N_4 - 
\tfrac{916241}{14515200} \, N_3 - \tfrac{87}{1280} \, N_2 - \tfrac{71941}{403200} \, N_1 + \tfrac{487467767}{3135283200} \, \pi^2 
- \tfrac{3579168251}{1306368000} \, \pi^4 \nonumber \\ &&- \tfrac{36666499}{228614400} \, 
\pi^6 - \tfrac{224}{15} \, a_5 - \tfrac{313387}{16200} \, a_4 - \tfrac{87}{2240} \, \zeta_7 + \tfrac{10395883727}{54432000} \, \zeta_5 
- \tfrac{29}{1600} \, \zeta_5 \, \pi^2 + \tfrac{40758646507}{130636800} \, \zeta_3 + \tfrac{463655219}{97977600} \, \zeta_3 \, \pi^2 \nonumber \\
&&- \tfrac{29}{4800} \, \zeta_3 \, \pi^4 + \tfrac{15591581}{2721600} \, \zeta_3^2 - \tfrac{119}{675} \, 
l_2 \, \pi^4 + \tfrac{313387}{388800} \, l_2^2 \, \pi^2 - \tfrac{28}{135} \, l_2^3 \, \pi^2 - \tfrac{313387}{388800} \, 
l_2^4 + \tfrac{28}{225} \, l_2^5 + \tfrac{4315007269}{4976640} \, l_m + \tfrac{103}{72} \, l_m \, N_3 \nonumber \\ &&+ \tfrac{87}{320} \, l_m
\, N_1 - \tfrac{5492587}{414720} \, l_m \, \pi^2 + \tfrac{1313957}{518400} \, l_m \, \pi^4 - \tfrac{29}{10080} \, l_m \, 
\pi^6 + \tfrac{448}{45} \, l_m \, a_4 - \tfrac{2928779}{21600} \, l_m \, \zeta_5 - \tfrac{32483549}{103680} \, l_m \, \zeta_3 \nonumber \\ &&- 
\tfrac{61259}{38880} \, l_m \, \zeta_3 \, \pi^2 - \tfrac{29}{120} \, l_m \, \zeta_3^2 - \tfrac{56}{135} \, l_m \, l_2^2 \, \pi^2 + \tfrac{56}{135} \, l_m \, l_2^4 
+ \tfrac{259}{108} \, l_m^2 - \tfrac{11}{10} \, l_m^3 \biggr)
\nonumber \\
&+& C_F \, C_A^2  \,  \biggl( \tfrac{25490069288387}{112870195200} - \tfrac{227}{2835} \, N_5 + \tfrac{1199}{8640} \, N_4
- \tfrac{413339}{9676800} \, N_3 + \tfrac{91}{3840} \, N_2 + \tfrac{3511}{37800} \, N_1 + \tfrac{1042284611}{9405849600} \, \pi^2 
+ \tfrac{361381739}{1306368000} \, \pi^4 \nonumber \\ &&+ \tfrac{42598397}{685843200} \, \pi^6
- \tfrac{1712}{45} \, a_5 + \tfrac{272567}{10800} \, a_4 + \tfrac{13}{960} \, \zeta_7 - \tfrac{613974553}{9072000} \, \zeta_5 + 
\tfrac{91}{14400} \, \zeta_5 \, \pi^2 + \tfrac{277861513}{195955200} \, \zeta_3 - \tfrac{15355283}{8164800} \, \zeta_3 \, \pi^2
+ \tfrac{91}{43200} \, \zeta_3 \, \pi^4 \nonumber \\ &&- \tfrac{18234067}{8164800} \, \zeta_3^2 - \tfrac{1819}{4050} \, l_2 \, \pi^4 - 
\tfrac{272567}{259200} \, l_2^2 \, \pi^2 - \tfrac{214}{405} \, l_2^3 \, \pi^2 + \tfrac{272567}{259200} \, l_2^4 + 
\tfrac{214}{675} \, l_2^5 - \tfrac{15657616211}{44789760} \, l_m - \tfrac{1199}{2160} \, l_m \, N_3 - \tfrac{91}{960} \, l_m \, 
N_1 \nonumber \\ &&+ \tfrac{3652601}{746496} \, l_m \, \pi^2 - \tfrac{2078929}{1555200} \, l_m \, \pi^4 + \tfrac{13}{12960} \, l_m \, 
\pi^6 + \tfrac{3424}{135} \, l_m \, a_4 + \tfrac{3411823}{64800} \, l_m \, \zeta_5 + \tfrac{42077383}{311040} \, l_m \, \zeta_3
+ \tfrac{73807}{116640} \, l_m \, \zeta_3 \, \pi^2 \nonumber \\ &&+ \tfrac{91}{1080} \, l_m \, \zeta_3^2 - \tfrac{428}{405} \, l_m \, l_2^2 \, \pi^2
+ \tfrac{428}{405} \, l_m \, l_2^4 - \tfrac{13597}{14580} \, l_m^2 + \tfrac{121}{270} \, l_m^3 \biggr),
\ea
\end{widetext}
where $C_F$ and $C_A$ are Casimir operators of the fundamental
and adjoint representations respectively. In the case of the SU(N)
group $C_F = (N^2-1)/2N$ and $C_A = N$. Furthermore, $l_m =
\log(m^2(\mu)/\mu^2)$, $l_2 = \log 2$, $\zeta_i$ are Riemann 
zeta numbers, and $a_i = {\rm Li}_i(1/2)$ with ${\rm Li}_i(x)$
the polylogarithm function. The numerical constants $N_i$ are
defined by the $\epsilon$ expansion of the vacuum integrals given in
Fig.~\ref{masters}, with the understanding that the integration
measure per loop is $e^{\epsilon \gamma_E}/i\pi^{2-\epsilon}\int d^d k$
\ba
\label{num}
N_1 &=& + 2369.669517745751, \nonumber \\
N_2 &=& + 9090.208679977050, \nonumber \\
N_3 &=& - 764.0948373358558, \nonumber \\
N_4 &=& - 4647.352454831194, \nonumber \\
N_5 &=& - 1.808879546208335.
\ea
The above numbers have been obtained by changing the normalization of the
high precision results of \cite{Schroder:2005va}. In fact, we took the
values of all the master integrals needed from there \footnote{Note
that there is a misprint in Eq. 6.38 of \cite{Schroder:2005va}, the
term $+\pi^4/20\; \epsilon^4$ should read $-\pi^4/20\; \epsilon^4$ (the
original source of this result, \cite{Laporta:2002pg}, is
correct). Similarly, the term $+36\; \zeta_3\; \epsilon^5$ in Eq. 6.40,
should read $-36\; \zeta_3\; \epsilon^5$.}. Relations between the
numbers in Eq.~(\ref{num}) discovered in \cite{Chetyrkin:2006dh} have
not been used in our analysis. We checked, however, the values of all
integrals up to six lines with the help of Mellin-Barnes integral
representations and the package \cite{Czakon:2005rk}.

The single fermion loop contribution to $\overline{\rm C}_1^{(3)}$
with $n_l$ massless quark species is given by
\begin{widetext}
\ba
&+& C_F^2 \, T_F  \,  \biggl(  - \tfrac{155640216647897}{526727577600} + \tfrac{30661}{322560} \, N_1 - 
\tfrac{391510309}{179159040} \, \pi^2 - \tfrac{288518747}{174182400} \, \pi^4 - \tfrac{30661}{30481920} \, 
\pi^6 + \tfrac{127252}{945} \, a_4 + \tfrac{30661}{89600} \, \zeta_5 \nonumber \\ &&+ \tfrac{318864713617}{1828915200} \, \zeta_3 + 
\tfrac{30661}{161280} \, \zeta_3 \, \pi^2 - \tfrac{30661}{362880} \, \zeta_3^2 - \tfrac{31813}{5670} \, l_2^2 \, \pi^2 + 
\tfrac{31813}{5670} \, l_2^4 - \tfrac{228583}{19440} \, l_m + \tfrac{139087}{12960} \, l_m \, \zeta_3 - \tfrac{22}{135} \, l_m^2
+ \tfrac{2}{5} \, l_m^3 \biggr)
\nonumber \\
&+& C_F^2 \, T_F \, n_l  \,  \biggl( \tfrac{1080947}{19440} + \tfrac{4793}{103680} \, N_3 - \tfrac{52723}{233280} \, \pi^2 + 
\tfrac{117811}{194400} \, \pi^4 - \tfrac{4793}{270} \, a_4 - \tfrac{857947}{194400} \, \zeta_5 - \tfrac{25009261}{466560}
\, \zeta_3 - \tfrac{4793}{69984} \, \zeta_3 \, \pi^2 + \tfrac{4793}{6480} \, l_2^2 \, \pi^2 \nonumber \\ &&- \tfrac{4793}{6480} \, l_2^4 - 
\tfrac{19801}{2160} \, l_m + \tfrac{120817}{12960} \, l_m \, \zeta_3 - \tfrac{22}{135} \, l_m^2 + \tfrac{2}{5} \, l_m^3 \biggr)
\nonumber \\
&+& C_F \, C_A \, T_F  \,  \biggl(  - \tfrac{11189587381799}{429185433600} + \tfrac{65}{2592} \, N_3 + \tfrac{11339}{645120} \, N_1 
- \tfrac{188717291}{358318080} \, \pi^2 - \tfrac{6364753}{69672960} \, \pi^4 - \tfrac{11339}{60963840} \, \pi^6 
+ \tfrac{8194}{945} \, a_4 + \tfrac{1724977}{43545600} \, \zeta_5 \nonumber \\ &&+ \tfrac{360428176373}{40236134400} \, \zeta_3 
- \tfrac{156623}{78382080} \, \zeta_3 \, \pi^2 - \tfrac{11339}{725760} \, \zeta_3^2 - \tfrac{4097}{11340} \, l_2^2 \, \pi^2 
+ \tfrac{4097}{11340} \, l_2^4 - \tfrac{775451}{349920} \, l_m - \tfrac{7013}{25920} \, l_m \, 
\zeta_3 + \tfrac{1051}{3645} \, l_m^2 - \tfrac{44}{135} \, l_m^3 \biggr)
\nonumber \\
&+& C_F \, C_A \, T_F \, n_l  \,  \biggl(  - \tfrac{31394101}{1049760} - \tfrac{4793}{207360} \, N_3 + \tfrac{52723}{466560} \, 
\pi^2 - \tfrac{26483}{388800} \, \pi^4 + \tfrac{4793}{540} \, a_4 + \tfrac{857947}{388800} \, \zeta_5 + \tfrac{7792237}{933120} \, \zeta_3 
+ \tfrac{4793}{139968} \, \zeta_3 \, \pi^2 \nonumber \\ &&- \tfrac{4793}{12960} \, l_2^2 \, \pi^2 + \tfrac{4793}{12960} \, 
l_2^4 - \tfrac{627893}{116640} \, l_m + \tfrac{4183}{2880} \, l_m \, \zeta_3 + \tfrac{1051}{3645} \, l_m^2 - \tfrac{44}{135} \, 
l_m^3 \biggr),
\ea
\end{widetext}
where $T_F$ is the trace of the fundamental representation, which we
take to have the standard value $T_F = 1/2$. Our result does not
depend on the number of heavy quark species of mass $m$. It should be
understood, however, that all terms without $n_l$ come from the heavy
quark loops.

The double fermion loop contribution to $\overline{\rm C}_1^{(3)}$ has
been published before in \cite{Chetyrkin:2004fq}. We found complete
agreement with that result, and reproduce it here only for completeness
\begin{widetext}
\ba
&& C_F \, T_F^2  \,  \biggl( \tfrac{163868}{98415} - \tfrac{3287}{2430} \, \zeta_3 - \tfrac{14483}{21870} \, l_m + \tfrac{203}{324} \, l_m
\, \zeta_3 + \tfrac{236}{3645} \, l_m^2 + \tfrac{8}{135} \, l_m^3 \biggr)
+ C_F \, T_F^2 \, n_l  \,  \biggl(  - \tfrac{782857}{65610} - \tfrac{29}{2592} \, N_3 + \tfrac{319}{5832} \, \pi^2 \nonumber \\ &&- \tfrac{319}{4860} \, \pi^4 
+ \tfrac{116}{27} \, a_4 + \tfrac{5191}{4860} \, \zeta_5 + \tfrac{340313}{58320} \, \zeta_3 + \tfrac{145}{8748} \, \zeta_3 \, 
\pi^2 - \tfrac{29}{162} \, l_2^2 \, \pi^2 + \tfrac{29}{162} \, l_2^4 - \tfrac{3779}{21870} \, l_m + \tfrac{203}{324} \, l_m \, \zeta_3
+ \tfrac{472}{3645} \, l_m^2 + \tfrac{16}{135} \, l_m^3 \biggr)
\nonumber \\
&&+ C_F \, T_F^2 \, n_l^2  \,  \biggl( \tfrac{42173}{32805} - \tfrac{112}{135} \, \zeta_3 + \tfrac{1784}{3645} \, l_m + \tfrac{236}{3645} \, 
l_m^2 + \tfrac{8}{135} \, l_m^3 \biggr).
\ea
\end{widetext}

Substituting the color factors and the numerical values of the
constants, our complete result can be cast in the following form
\ba
&&1.87882 - 2.79472 \,n_l + 0.0961014 \,n_l^2 \\ &&+
        18.4050 \,l_m - 5.26881 \,n_l \,l_m +
        0.163146 \,n_l^2 \,l_m \nonumber \\ && - 0.660174 \,l_m^2 +
        0.474989 \,n_l \,l_m^2 +
        0.0215821 \,n_l^2 \,l_m^2 \nonumber\\ && +
        0.656790 \,l_m^3 - 0.256790 \,n_l \,l_m^3 +
        0.0197531 \,n_l^2 \,l_m^3. \nonumber
\ea

In order to determine the mass of a given quark from our perturbative
result we shall use the moments of the hadronic cross section
defined by
\be
\label{exp}
M_n = \int \frac{ds}{s^{n+1}} R(s).
\ee
With the help of a dispersion relation, this can be translated into
\be
M_n = \frac{12 \pi^2}{n!}\left(\frac{d}{dq^2}\right)^n \Pi_h(q^2)\; \big|_{q^2=0}.
\ee
If one compares the above with Eq.~(\ref{moments}), then the mass is
finally given by
\be
m(\mu) = \frac{1}{2} \left( \frac{9}{4}Q_h^2 \;\frac{\overline{C}_n}{M_n^{\rm exp}}\right)^{1/(2n)}.
\ee
As explained at the beginning, we are only interested in the first
moment. The experimental value, Eq.~(\ref{exp}), is made out of three constituents, the
resonances, the threshold region and the continuum. We note that tiny
changes in the central value of $\alpha_s(M_Z)$ \cite{Bethke:2004uy}
entering trough the perturbative result used for the continuum
have no impact on the final moments. We observe the same with respect
to parts of the five-loop result \cite{Baikov:2004ku}. Therefore,
apart from the resonance data for $J/\psi$ and $\psi'$
\cite{Eidelman:2004wy}, we adopt the moments from
\cite{Kuhn:2001dm}. Note that an alternative set of experimental
moments that would lead to a larger error estimate can be found in
\cite{Corcella:2002uu,Hoang:2004xm}. After correction for the
resonances, we have
\ba
\left(M^{\rm exp}_1\right)_{\rm charm} &=& 0.2087 \pm 0.0042, \\
\left(M^{\rm exp}_1\right)_{\rm bottom} &=& 0.004456 \pm 0.000121.
\ea
Similarly to \cite{Kuhn:2001dm}, we first determine the mass of the
charm quark at 3~GeV, and then run it down to
$\overline{m}_c(\overline{m}_c)$. For consistency reasons the running
is performed with three-loop accuracy, {\it i.e.} to ${\cal
  O}(\alpha_s^3)$. We use the package \cite{Chetyrkin:2000yt} and
obtain
\be
\overline{m}_c(\overline{m}_c) = 1.295 \pm 0.009_{\alpha_s} \pm
0.0003_{\mu} \pm 0.012_{\rm exp.}\; {\rm GeV}.
\ee
The subscripts denote the various sources of the error (variation of
$\alpha_s$ within error bars, variation of $\mu$ in the range $3 \pm
1$ GeV and variation of the first moment). The combined error given
in the abstract is obtained by adding the errors in squares. We note
that the four-loop result gives a shift of -2~MeV whereas the new
resonance data -10~MeV. Finally, we observe an almost threefold reduction of
the perturbative error.

Performing a similar analysis for the b-quark, where the calculation
is first done at 10~GeV and then the mass is run down to
$\overline{m}_b(\overline{m}_b)$, we obtain
\be
\overline{m}_b(\overline{m}_b) = 4.205 \pm 0.010_{\alpha_s} \pm
0.002_{\mu} \pm 0.057_{\rm exp.} \; {\rm GeV}.
\ee
The only change in the error determination with respect to the case of
the c-quark is connected to the
variation of $\mu$ in the range $10 \pm 5$ GeV. We note, that our
result is exactly the same as that of the three-loop analysis in
\cite{Kuhn:2001dm}, apart from a reduction of the perturbative error
by almost an order of magnitude. We would observe a shift of the
central value, if the caculation were performed starting from low
values of $\mu$ close to 5~GeV. This is contained in our error
estimate from the variation of $\mu$.
As a final remark, let us mention that there is some discussion on the resonance
contribution of $\Upsilon(4S)$ and $\Upsilon(5S)$
\cite{Corcella:2002uu}. Adopting the values suggested by the latter paper, we
observe a small shift of a few MeV in the central value, well within
our error estimate.

The authors would like to thank J. K\"uhn for an interesting
discussion on quark masses and sum rules. Parts of the presented
calculations were performed on the DESY Zeuthen Grid Engine computer cluster.
This work was supported by the Sofja Kovalevskaja Award of the
Alexander von Humboldt Foundation sponsored by the German Federal
Ministry of Education and Research.


\begin{thebibliography}{00}

\bibitem{Dougall:2005ev}
A.~Dougall, C.~M.~Maynard and C.~McNeile,
JHEP {\bf 0601}, 171 (2006);
C.~McNeile, C.~Michael and G.~Thompson  [UKQCD Collaboration],
Phys.\ Lett.\ B {\bf 600}, 77 (2004).

\bibitem{Penarrocha:2001ig}
J.~Penarrocha and K.~Schilcher,
Phys.\ Lett.\ B {\bf 515}, 291 (2001);
J.~Bordes, J.~Penarrocha and K.~Schilcher,
Phys.\ Lett.\ B {\bf 562}, 81 (2003).

\bibitem{Eidemuller:2000rc}
M.~Eidemuller and M.~Jamin,
Phys.\ Lett.\ B {\bf 498}, 203 (2001).

\bibitem{Hoang:2004wh}
A.~H.~Hoang,
arXiv:hep-ph/0412160.

\bibitem{Kuhn:2001dm}
J.~H.~K\"uhn and M.~Steinhauser,
Nucl.\ Phys.\ B {\bf 619}, 588 (2001)
[Erratum-ibid.\ B {\bf 640}, 415 (2002)].

\bibitem{Corcella:2002uu}
G.~Corcella and A.~H.~Hoang,
Phys.\ Lett.\ B {\bf 554} (2003) 133.

\bibitem{Hoang:2004xm}
A.~H.~Hoang and M.~Jamin,
Phys.\ Lett.\ B {\bf 594} (2004) 127.

\bibitem{Chetyrkin:2006xg}
K.~G.~Chetyrkin, J.~H.~K\"uhn and C.~Sturm,
arXiv:hep-ph/0604234.

\bibitem{Eidelman:2004wy}
S.~Eidelman {\it et al.}  [Particle Data Group Collaboration],
Phys.\ Lett.\ B {\bf 592}, 1 (2004).

\bibitem{Novikov:1977dq}
V.~A.~Novikov, L.~B.~Okun, M.~A.~Shifman, A.~I.~Vainshtein, M.~B.~Voloshin and V.~I.~Zakharov,
Phys.\ Rept.\  {\bf 41}, 1 (1978).

\bibitem{Reinders:1984sr}
L.~J.~Reinders, H.~Rubinstein and S.~Yazaki,
Phys.\ Rept.\  {\bf 127}, 1 (1985).

\bibitem{Shifman:1978bx}
M.~A.~Shifman, A.~I.~Vainshtein and V.~I.~Zakharov,
Nucl.\ Phys.\ B {\bf 147}, 385 (1979);
Nucl.\ Phys.\ B {\bf 147}, 448 (1979).

\bibitem{Chetyrkin:1995ii}
K.~G.~Chetyrkin, J.~H.~K\"uhn and M.~Steinhauser,
Phys.\ Lett.\ B {\bf 371}, 93 (1996);
Nucl.\ Phys.\ B {\bf 505}, 40 (1997).

\bibitem{Bai:2001ct}
J.~Z.~Bai {\it et al.}  [BES Collaboration],
Phys.\ Rev.\ Lett.\  {\bf 88}, 101802 (2002)
[arXiv:hep-ex/0102003].

\bibitem{Groote:2001py}
JETP Lett.\  {\bf 75}, 221 (2002)
[Pisma Zh.\ Eksp.\ Teor.\ Fiz.\  {\bf 75}, 267 (2002)];
Eur.\ Phys.\ J.\ C {\bf 21}, 133 (2001).

\bibitem{Portoles:2002rt}
J.~Portoles and P.~D.~Ruiz-Femenia,
Eur.\ Phys.\ J.\ C {\bf 24}, 439 (2002).

\bibitem{Broadhurst:1993mw}
D.~J.~Broadhurst, J.~Fleischer and O.~V.~Tarasov,
Z.\ Phys.\ C {\bf 60}, 287 (1993)
[arXiv:hep-ph/9304303].

\bibitem{barbieri}
R.~Barbieri and E.~Remiddi, Nuovo Cim.{\bf 13A}, 99 (1973).

\bibitem{Laporta:2001dd}
S.~Laporta,
Int.\ J.\ Mod.\ Phys.\ A {\bf 15}, 5087 (2000)

\bibitem{Czakon:2004bu}
M.~Czakon,
Nucl.\ Phys.\ B {\bf 710}, 485 (2005).

\bibitem{Schroder:2005va}
Y.~Schroder and A.~Vuorinen,
JHEP {\bf 0506}, 051 (2005).

\bibitem{Laporta:2002pg}
S.~Laporta,
Phys.\ Lett.\ B {\bf 549}, 115 (2002);
S.~Laporta and E.~Remiddi,
Phys.\ Lett.\ B {\bf 379}, 283 (1996).

\bibitem{Chetyrkin:2006dh}
K.~G.~Chetyrkin, M.~Faisst, C.~Sturm and M.~Tentyukov,
arXiv:hep-ph/0601165.

\bibitem{Czakon:2005rk}
M.~Czakon,
arXiv:hep-ph/0511200.

\bibitem{Chetyrkin:2004fq}
K.~G.~Chetyrkin, J.~H.~K\"uhn, P.~Mastrolia and C.~Sturm,
Eur.\ Phys.\ J.\ C {\bf 40}, 361 (2005).

\bibitem{Bethke:2004uy}
S.~Bethke,
Nucl.\ Phys.\ Proc.\ Suppl.\  {\bf 135}, 345 (2004)
[arXiv:hep-ex/0407021].

\bibitem{Baikov:2004ku}
P.~A.~Baikov, K.~G.~Chetyrkin and J.~H.~K\"uhn,
Nucl.\ Phys.\ Proc.\ Suppl.\  {\bf 135}, 243 (2004).

\bibitem{Chetyrkin:2000yt}
K.~G.~Chetyrkin, J.~H.~K\"uhn and M.~Steinhauser,
Comput.\ Phys.\ Commun.\  {\bf 133}, 43 (2000).

\end{thebibliography}
\end{document}